\documentclass[sigconf]{acmart}

\usepackage{booktabs} 
\usepackage{caption}
\usepackage{subfig}
\usepackage{url}
\usepackage{epstopdf}
\usepackage{array}
\usepackage{pifont}
\usepackage{multirow}
\usepackage{amssymb}
\usepackage{footnote}
\usepackage{tablefootnote}

\acmDOI{}

\acmISBN{}

\acmConference[SMERP'18]{SMERP workshop with The Web Conference}{2018}{} 
\acmYear{2018}
\copyrightyear{2018}

\acmPrice{15.00}

\usepackage{color}

\begin{document}

\title{SAVITR: A System for Real-time Location Extraction from Microblogs during Emergencies}

\author{Ritam Dutt, Kaustubh Hiware, Avijit Ghosh, Rameshwar Bhaskaran}
\affiliation{\institution{Indian Institute of Technology Kharagpur, India}}


\begin{abstract}
We present SAVITR, a system that leverages the information posted on the Twitter microblogging site to monitor and analyse emergency situations.
Given that only a very small percentage of microblogs are geo-tagged, it is essential for such a system to extract locations from the text of the microblogs.
We employ natural language processing techniques to infer the locations mentioned in the microblog text, in an unsupervised fashion and display it on a map-based interface. 
The system is designed for efficient performance, achieving an F-score of 0.79, and is approximately two orders of magnitude faster than other available tools for location extraction.

\end{abstract}

%
%

\ccsdesc[500]{Information systems~Information retrieval}



\keywords{Emergencies, microblogs, location extraction, Geonames}

\maketitle

\section{Introduction}

Online social media sites, especially microblogging sites like Twitter and Weibo, have been shown to be very useful for gathering  situational information in real-time~\cite{social-media-emergency-survey,rudra-cikm-disaster}. 
Consequently, it is imperative to not only process the vast incoming data stream on a real-time basis, but also to extract relevant information  from the unstructured and noisy data accurately.

It is especially crucial to extract geographical locations from tweets (microblogs), since the locations help to associate the information available online with the physical locations.
This task is challenging since geo-tagged tweets are very sparse, especially in developing countries like
India, accounting for only 0.36\% of the total tweet traffic.
Hence it becomes necessary to extract locations from the text of the tweets.

This work proposes a novel and fast method of extracting locations from English tweets posted during emergency situations. 
The location is inferred from the tweet-text in an unsupervised fashion as opposed to using the geo-tagged field. 
Note that several methodologies for extracting locations from tweets
have been proposed in literature; some of these are discussed in the next section. 
We compare the proposed methodology with several existing methodologies
in terms of coverage (Recall) and accuracy (Precision). 
Additionally, we also compared the speed of operation of different methods, which is crucial for real-time deployment of the methods. 
The proposed method achieves very competitive values of Recall and Precision with the baseline methods, and the highest F-score among all methods.
Importantly, the proposed methodology is several orders of magnitude faster than most of the prior methods, and is hence 
suitable for real-time deployment. 

We deploy the proposed methodology on a system available at \url{http://savitr.herokuapp.com}, which is described in a later section.

\section{Related Work}

We discuss some existing information systems for use during emergencies, and some prior methods for location extraction from microblogs.

\subsection{Information Systems}

A few Information Systems have already been implemented in various countries for emergency informatics, and their efficacy has been demonstrated in a variety of situations. 
Previous work on real-time earthquake detection in Japan was deployed by~\cite{Sakaki} using Twitter users as social sensors. 
Simple systems like the Chennai Flood Map~\cite{Mapbox}, which combines crowdsourcing and open source mapping,  have demonstrated the need and utility of Information Systems during the 2015 Chennai floods. 
Likewise, Ushahidi~\cite{Ushahidi} enables local observers to submit reports using their mobile phones or the Internet, thereby creating a temporal and geospatial archive of an ongoing event. 
Ushahidi has been deployed in situations such as earthquakes in Haiti, Chile, forest fires in Italy and Russia.

Our system also works on the same basic principle as the aforementioned ones -- information extraction from crowdsourced data. 
However, unlike Mapbox~\cite{Mapbox} and Ushahidi~\cite{Ushahidi}, it is not necessary for the users to explicitly specify the location. Rather, we infer it from the tweet text, without any prior manual labeling.

\subsection{Location Inferencing methods}

Location inferencing is a specific variety of Named Entity Recognition (NER), whereby only the entities corresponding to valid geographical locations are extracted. 
There have been seminal works regarding location extraction from microblog text, inferring the location of a user from the user's set of posted tweets and even predicting the probable location of a tweet by training on previous tweets having valid geo-tagged fields.
Publicly available tools like Stanford NER \cite{StanfordNER}, TwitterNLP \cite{TwitterNLP1}, OpenNLP \cite{OpenNLP} and Google Cloud\footnote{https://cloud.google.com/natural-language/}, are also available for tasks such as location extraction from text.

We focus our work only on extracting the locations from the tweet text, since we have observed that 
(i)~a very small fraction of tweets are geo-tagged, and 
(ii)~even for geo-tagged tweets, a tweet's geo-tagged location is not always a valid representative of the incident mentioned in the tweet text. 
For instance, the tweet ``{\it Will discuss on \@TimesNow at 8.30 am today regarding Dengue Fever in Tamil Nadu.}'' clearly refers to Tamil Nadu, 
but the geo-tagged location is New Delhi (from where the tweet was posted).

We give an overview of the different types of methodologies used
in location extraction systems.
Prior state-of-the-art methods have performed common preprocessing steps like noun-phrase extraction and phrase matching~\cite{Malmasi}, or regex matching~\cite{twitie} before employing the following techniques for location extraction. 
\begin{itemize}
\item{Gazetteer lookup:}
Gazetteer based search and n-gram based matching have been employed by  \cite{Malmasi}, \cite{middleton} , \cite{gelernter}. Usually some publicly available gazetteers like GeoNames or OpenStreetMap are used.

\item{Handcrafted rules} were employed in \cite{Malmasi} and \cite{gelernter}

\item{Supervised methods:}
Well-known supervised models used in this current context are:
\begin{enumerate}
\item{Models based on Conditional Random Fields (CRF)} such as the Stanford NER parser which was employed by \cite{gelernter} and \cite{Malmasi}. While \cite{gelernter} trained the model on tweet texts, \cite{Malmasi} used the parser without training.
\item{Maximum entropy based models} such as the OpenNLP was deployed by \cite{lingad} without training and it infers location using ME.
\end{enumerate}

\item{Semi-supervised methods:}
The work~\cite{ji2016joint} used semi-supervised methods such as beam-search and structured perceptrons to label sequences and linked them with corresponding Foursquare location entities. 

\end{itemize}


\section{Extracting locations from microblogs}

We now describe the proposed methodology for inferring locations from tweet text. The methodology involves the following tasks.

\subsection{Hashtag Segmentation}

Hashtags are a relevant source of information in Twitter. Especially for tweets posted during emergency situations, hashtags often contain location names embedded in them, e.g., \#NepalQuake, \#GujaratFloods.
However, due to the peculiar style of coining hashtags, it becomes imperative to break them into meaningful words. 
Similar to~\cite{Malmasi} and~\cite{DBLP:journals/corr/abs-1708-03105}, we adopt a statistical word segmentation based algorithm~\cite{Peter} to break a hashtag into distinct words, and extract locations from the distinct words. 
We also retain the original hashtag, to ensure we do not lose out on meaningful remote locations simply because they are uncommon. 




We have observed that hashtag segmentation has some unforeseen outcomes. While trying to optimize recall from a tweet, it hampers precision, especially when the segmented words corresponds to actual locations. 
For example `\#Bengaluru' (a place in India) is broken down into `bengal' and `uru', which are two other places in India. 
Again `\#Kisiizi' (name of a hospital in Uganda) is incorrectly segmented into `kissi' and `zi', none of which are location names.

In spite of these limitations of hashtag segmentation, we still carry out this step since we seek to extract all possible location names, including those embedded within hashtags.

\subsection{Tweet Preprocessing}

We then applied common pre-processing techniques on the tweet text and removed URLs, mentions, and stray characters like 'RT', brackets, \# and ellipses and segmented CamelCase words. We did not perform case-folding on the text since we wanted to detect proper nouns. 
Likewise, we also abstained from stemming since location names might get altered and cannot be detected using the gazetteer.

\subsection{Disambiguating Proper Nouns from Parse Trees}

Since most location names are likely to be proper nouns, we use a heuristic to determine whether a proper noun is a location. 
We first apply a Parts-of-Speech (POS) tagger to yield POS tags.
There are several POS taggers publicly available, which could be applied, such as SPaCy\footnote{https://spacy.io/}, the Twitter-specific 
CMU TweeboParser\footnote{http://www.cs.cmu.edu/~ark/TweetNLP/}, and so on.
We employ the POS Tagger of SPaCy, in preference to the CMU TweeboParser, due to the heavy processing time of the latter. The TweeboParser was 1000 times slower as opposed to SpaCy. 
We considered the speed to be a viable trade-off for accuracy since we want the method to be deployed on a real-time basis and we observed the processing time would be a bottleneck in this regard.

Let $T_{i}$ denote the POS tag of the $i^{th}$ word $w_i$ of the tweet. If $T_{i}$ corresponds to a proper noun, we keep on appending words that succeed $w_i$, provided they are also proper nouns, delimiters or adjectives. 
We develop a list of common suffixes of location names (explained below). 
If $w_i$ is followed by a noun in this suffix list, we consider it to be a viable location. Acknowledging the fact that Out of Vocabulary (OOV) words are common in Twitter, we also consider those words which have a high Jaro-Winkler similarity with the words in the suffix list.

We also check the word immediately preceding $w_i$, to see if it is a preposition that usually precedes a place or location, such as `at', `in', `from', `to', `near', etc. 
We then split the stream of words via the delimiters. Thus we attempt to infer from the text proper nouns which conform to locations from their syntactic structure.

\subsection{Regex matches}

As mentioned in the previous section, we have compiled a suffix list 
containing words that usually come after a location name. 
The suffix list comprises different naming conventions for
landforms\footnote{https://en.wikipedia.org/wiki/List\_of\_landforms}, 
roads\footnote{https://wiki.waze.com/wiki/India/Editing/Roads} \footnote{http://www.haringey.gov.uk}, 
buildings\footnote{https://en.wikipedia.org/wiki/List\_of\_building\_types} and towns.
A part of the suffix list is shown in Table~\ref{tab:Example list}.

We perform this additional task of regex similarity to account for cases when the tweet is posted in lowercase, making it difficult to detect and disambiguate proper nouns.
Using the suffix list enables us 
to detect places like `Vinayak hospital' and `Gujranwala town' from the tweet ``\textit{Urgent B+ group platelets suffering from dengue,Ankit Arora  At Vinayak hospital, Gujranwala town,delhi}''.

\begin{table}[tb]
\centering
	\begin{tabular}{|c|c|}
	\hline
	Type & Common Examples\\ \hline
	
	Landforms & doab, lake, steam, river, island, valley, mountain, hill \\
	Roads & street, st, boulevard, junction, lane, rd, avenue, bridge\\
	Buildings & hospital, school, shrine, cinema,villa, temple, mosque,  \\
	Towns & city, district, village, gram, place,town, nagar,  \\
	Directions & south, eastern, NW, SE, west, western, north east,\\
	\hline
	Diseases & dengue, ebola, cholera, zika, malaria, chikungunya  \\
	Disasters & earthquake, floods, drought, tsunami, landslide, rains\\
	
	\hline
	\end{tabular}
  	\caption{Examples of suffixes and emergency-related words}
	\label{tab:Example list}
\vspace*{-5mm}
\end{table}


\begin{figure}[tb]
	\centering
		\includegraphics[scale=0.23]{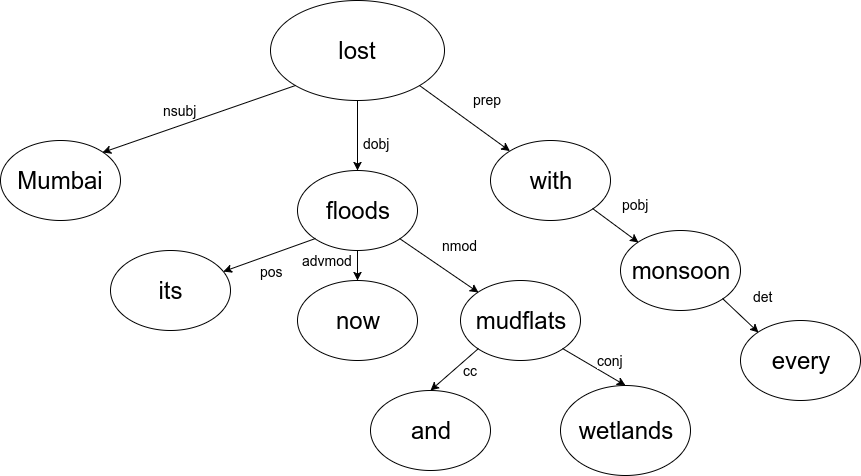}
		\caption{Dependency graph for a sample tweet ``{\it Mumbai lost its mudflats and wetlands, now floods with every monsoon.}''.}
		\label{fig:dependency_graph}
\vspace*{-5mm}
\end{figure}

\subsection{Dependency Parsing of Emergency words}

So far, the  methodology aims at improving the precision, but does not look to improve recall. This step is meant to improve recall by capturing 
those locations which do not follow the common patterns listed above.

Considering that our objective is to monitor emergency scenarios, we identify a set of words corresponding to 
epidemic disasters\footnote{https://en.wikipedia.org/wiki/List\_of\_epidemics} and natural disasters\footnote{https://en.wikipedia.org/wiki/Lists\_of\_disasters}, some of which are shown in Table ~\ref{tab:Example list}.
We identify the list of emergency words in the tweet text and consider words, namely proper nouns, nouns and adjectives, which are {\it at a short distance of 2-3 from the emergency word} in the dependency graph obtained for the tweet text. 
The distance metric refers to the number of links connecting the words in the dependency graph of the tweet text. A short dependency implies the word is more intimately affected by the emergency word. 

As an example, Figure~\ref{fig:dependency_graph} shows the dependency graph for the tweet ``{\it Mumbai lost its mudflats and wetlands, now floods with every monsoon.}''.
We see that the distance between Mumbai and floods in the dependency graph of the tweet is 2, whereas the actual distance between the words in the text is 7. Hence we can identify Mumbai as a proper location via dependency parsing,
Also, we extract the noun phrases from the dependency graph (as in~\cite{Malmasi}) and use the SpaCy NER tagger as in ~\cite{Malmasi,lingad,gelernter}.

\subsection{Gazetteer Verification}

The list of phrases and locations extracted by the above methods are then verified using a gazetteer, to retain only those words that correspond to real-world locations. For our system, the gazetteers also returns the geo-spatial coordinates to enable plotting the location on a map. 


\section{Comparative evaluation of the location inference}

In this section, we describe the evaluation of the proposed methodology,
and compare it with several baseline methods.
We start by describing the dataset and some design choices made by us.

\subsection{Dataset}

We used the Twitter Streaming API\footnote{https://developer.twitter.com/en/docs}, to collect tweets from 12$^{th}$ September, 2017 to 13$^{th}$ October, 2017, and filtered those tweets that contained either of the two words 'dengue' or 'flood'. 
This step produced a dataset of 317,567 tweets collected over a period of 31 days. The tweets were preprocessed to remove duplicates and also tweets written in non-English languages. 
This filtering resulted in 239,276 distinct tweets. 

\subsection{Gazetteer employed}

In this work, we currently focus on collecting and displaying tweets within the bounding box of the country of India. 
Thus, we need some lexicon / gazetteer to disambiguate whether a place is located inside India and what are its geographical coordinates. To that end, we scraped the data publicly available from Geonames\footnote{http://www.geonames.org/} and made a dictionary corresponding to different locations within India. The dictionary has the information of 449,973 locations within India. 
However, some places mentioned in this dictionary have high orthographic similarity with common English nouns. For example, we find that 
the word `song' is a place located in Sikkim, whose coordinates are \(27.24641 'N, 88.50622 'E\). Moreover, Geonames does {\it not} contain fine-grained information of addresses and places like roads and buildings. 

Consequently, we explored another gazetteer -- 
the Open Street Map gazetteer\footnote{http://geocoder.readthedocs.io/providers/OpenStreetMap.html} 
which has a comprehensive list of all possibles addresses for India. However, the sheer volume of data, $\approx 530$ times larger than Geonames, hampers performance in a real-time setting. 
Also, API calls takes considerable time as opposed to querying the Geonames gazetter\footnote{http://download.geofabrik.de/asia/india.html}.

Thus the choice of the gazetteer is governed by a trade-off
between recall and efficiency. 
We report performances using both gazetteers in this paper. Hence we consider two variants of the proposed methodology:
\begin{itemize}
\item GeoLoc- Our proposed methodology using Geonames as the gazetteer. 
\item OSMLoc- Our proposed methodology using Open Street Maps as the gazetter. 
\end{itemize}

\subsection{Baseline methodologies}

We compared the proposed approach of our algorithm with several baseline methodologies which are enlisted below: 
\begin{itemize}
\item UniLoc- Take all unigrams in the processed tweet text and infer if any of those correspond to a possible location (by referring to a gazetteer). 
\item BiLoc- Similar to UniLoc, except we consider both unigrams and bigrams in the tweet text.
\item StanfordNER - Employs the NER of coreNLP parser~\cite{StanfordNER}. 
\item TwitterNLP - Employ the NER of Twitter NLP parser developed by Ritter et al.~\cite{TwitterNLP1}
\item Google Cloud- Use the Google cloud API to infer locations.
\item SpaCyNER - Use the trained SpaCy NER tagger. 
\end{itemize}	
For all the baseline methods, the potential locations are checked using the GeoNames gazetteer.

\subsection{Evaluation Measures}

Given a tweet text, we wish to infer all possible locations contained in the tweet. Thus we should prefer a method which has higher recall. However, since we also aim to plot the location obtained from the tweet, the precision of our extracted locations also matters. Hence we apply the following measures.
\begin{equation}
Precision =\frac{\left | \mbox{Correct Locations} \bigcap \mbox{Retrieved Locations}  \right |}{ \mbox{Retrieved Locations}}
\end{equation}
\begin{equation}
Recall =\frac{\left | \mbox{Correct Locations} \bigcap \mbox{Retrieved Locations}  \right |}{ \mbox{Correct Locations}}
\end{equation}
where `Correct locations' is the set of locations actually mentioned in a tweet, as found by human annotators, and
`Retrieved locations' is the set of locations inferred by a certain methodology from the same tweet.
To get an idea of both precision and recall, we use F-score which is the harmonic mean of precision and recall.

Moreover, since we wish to deploy the system on a real-time basis, the evaluation time taken by a method is also a justifiable metric.

\subsection{Evaluation results}

We randomly selected 1,000 tweets from the collected set of tweets (as described earlier), and asked human annotators to identify those tweets which contain some location names.
The annotators identified a set of 101 tweets that contained at least one location name. Hence the comparative evaluation is carried out over
this set of 101 tweets.


\begin{table}[tb]
	\centering
		\begin{tabular}{|c|c|c|c|r|}
			\hline
			Method&	Precision&	Recall&	F-score&	Timing (in s)\\
			\hline
			UNILoc&	0.3848&	0.7852&	0.5165&	0.0553\\
			BILoc&	0.4025&	0.8590&	0.5482&	0.0624\\
			StanfordNER& 0.8145&	0.6778&	0.7399&	175.0124\\
			TwitterNLP&	0.6251&	0.5474&	0.5836&	28.0001\\
			GoogleCloud &	0.6131&	0.5311&	0.5692&	NA\\
            SpaCyNER & \textbf{0.9659}& 0.5862&0.7296 &1.0891\\
            \hline
			GeoLoc &	0.7485&	0.8389&\textbf{0.7911}&	1.1901\\
			OSMLoc & 0.3096& \textbf{0.9060}&0.46153 &711.5817\\

			\hline
			
			\hline
		\end{tabular}
		\caption{Evaluation Performance of the baseline methods and the proposed methods (two variants, one using GeoNames gazetteer, and the other using Open Street Maps gazetteer).}
		\label{tab:Evaluation table}
        \vspace*{-5mm}
\end{table}

Table~\ref{tab:Evaluation table} compares the performances of the baseline methods and the proposed method.
The last column shows the average time in seconds needed to process the $101$ tweets that we are using for evaluation.
We observe that GeoLoc performs the best in terms of F1 score as compared to all other methods. It also scores high on precision, ranking only third to StanfordNER and SpaCyNER. The high precision of SpaCyNer is counterbalanced by its very bad recall due to which it was hardly able to detect remote places like Mohali and May Hosp. from the tweet \footnote{Urgent B + blood needed for a crit dengue patient at May Hosp. , Mohali,(Chandigarh)}. Mohali is however, detected by our GeoLoc algorithm.

The slight decrease in precision is attributed to some common words like `song', `monsoon', `parole' being chosen as potential locations due to incorrect hashtag segmentation, and then the gazetter tagging these words as locations, since these are also names of certain places in India. 

It can also be seen that, the proposed method using GeoNames gazetteer is much faster than the other methods which achieve comparable performance (e.g., StanfordNER).

\noindent {\bf Choice of gazetteer:}
As stated earlier, the Geonames gazetteer lacks information of a granular level. Consequently specific places pertaining to hospitals and streets are often not recognized as valid locations. This hampers the recall of the system, e.g., the proposed methodology was unable to detect `star hospital' in the tweet ``{\it We need O-ve blood grup for 8 years boy suffering with dengue in star hospital in karimnagar , please Contact}.''

Open Street Map (OSM) is able to detect such specific locations and thus exhibits the highest recall amongst all other methods. However, using OSM has the side-effect of classifying many simple noun phrases as valid locations. For instance, 
`silicon city' is detected as a location in the tweet ``{\it @rajeev\_mp  seems its time to rename Bangalore as Floods city I/O silicon city.}'', since `silicon city' is judged a shortening for the entry `Concorde Silicon Valley, Electronics City Phase 1, Vittasandra, Bangalore'. 
As a result of such errors, the method using OSM has the lowest precision score amongst all the methods.

\noindent {\bf Performance over the entire dataset:}
From the entire set of 239,276 distinct tweets, only 
3,493 were geo-tagged, out of which 869 were from India (which corresponds to a minute $0.36\%$ of the entire dataset). 
The number of tweets which were successfully tagged from the entire dataset, using our proposed technique and Geonames was 68,793, which corresponds to approximately $26.15\%$. 
Hence the coverage is increased drastically.
We manually observed many of the inferred locations, and found a large fraction of the correct. The method could identify niche and remote places in India, like `Ghatkopar', `Guntur', `Pipra village' and `Kharagpur', besides metropolitan cities like `Delhi', `Kolkata' and `Mumbai'.

\section{SAVITR: Deploying the location inference method}

\begin{figure}[htb]
	\centering
		\includegraphics[scale=0.23]{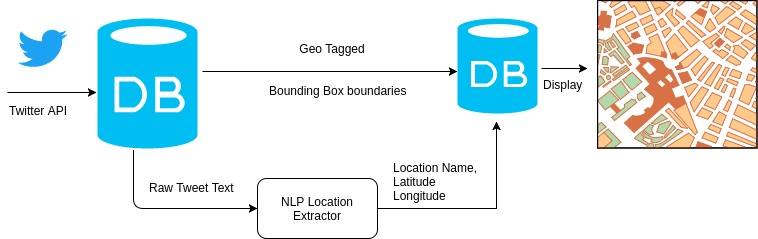}
		\caption{System architecture of the SAVITR system}
		\label{fig:system_arch}
        \vspace*{-5mm}
\end{figure}

\begin{figure}[tb]
	\centering
		\includegraphics[scale=0.23]{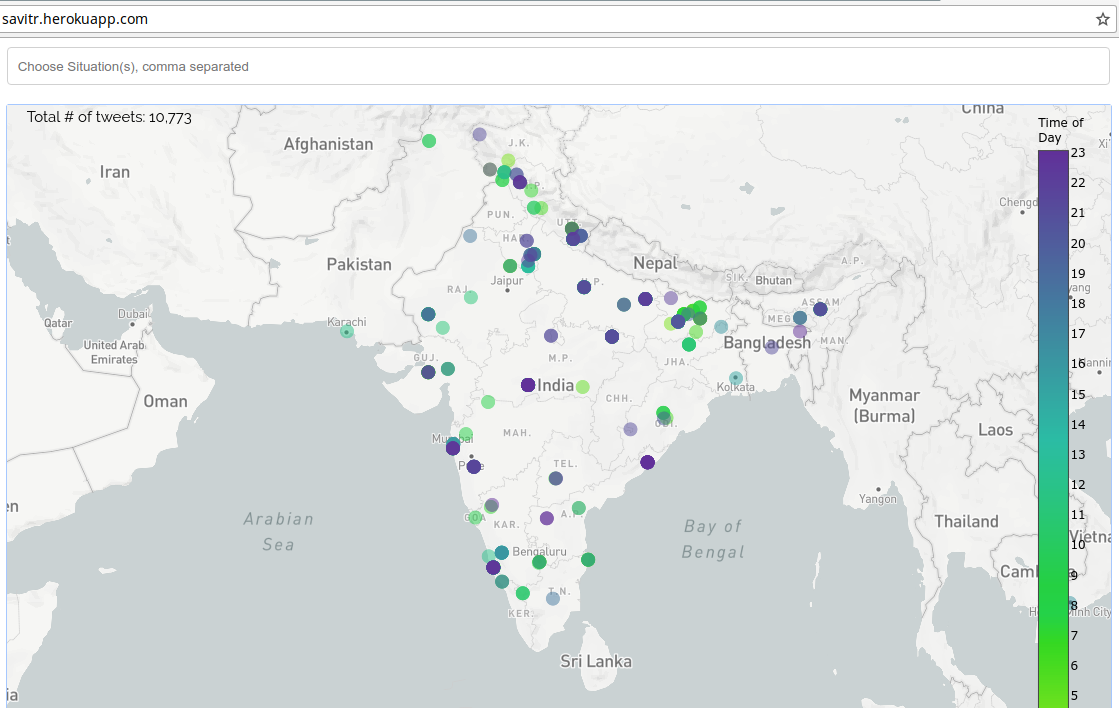}
		\caption{Snapshot of the SAVITR system: Tweets visualised on India's map}
		\label{fig:map_general}
	\vspace*{-5mm}
\end{figure}

We have deployed the proposed techniques (using GeoNames) on a system named SAVITR, which is live at \url{http://savitr.herokuapp.com}.
The software architecture of Savitr is presented in Figure~\ref{fig:system_arch}. 
Since the amount of data to be displayed is massive, we had to implement certain design considerations so that the information displayed is both compact and visually enriching, while at the same time scalable. 
The system was built using the Dash framework by Plotly~\cite{Plotly}. For our visualization purpose, we settled on a mapbox Map at the heart of the UI, with various controls, as described below. A snapshot
of the system is shown in Figure~\ref{fig:map_general}.
\begin{itemize}

\item A search bar at the top of the page. Whenever a term is entered into the search bar, the map refreshes and shows tweets pertaining to that query term. It also supports multiple search queries like "Dengue, Malaria".

\item The tweets on the map are color coded according to the time of the day. Tweets posted in the night are darker.

\item A date-picker -- if one wishes to visualize tweets posted during a particular time duration, this provides fine grained date selection, both at the month and date level.

\item A Histogram -- this shows the number of relevant (tagged) tweets posted per day.

\item Untagged tweets -- Finally, at the bottom of the page we display the tweets for which location could not be inferred (and hence they could not be shown on the map).
\end{itemize}

We report the performance of the system during the massive dengue outbreak that plagued India in the fall of 2017.\footnote{https://www.telegraphindia.com/india/dengue-spurt-in-south-182846} 
The state of Kerala was severely affected by the outbreak. 
During this period, as many as $2204$ tweets mentioning Kerala were identified by the system, which is far higher than the average rate at which `Kerala' is mentioned on any average day. 
Additionally, out of the $2204$ tweets containing the location `Kerala',
$1960$ ($88.92\%$) also contained the term `dengue' which is included in the list of disaster terms compiled by us (see Table~\ref{tab:Example list}).
These statistics demonstrate how the SAVITR system can be used as an `Early warning system' to flag any upcoming emergency situation.

Though the SAVITR system presently infers locations within India, it can be easily extended to infer locations within other countries, and the whole world in general. 


\section{Concluding discussion}

We proposed a methodology for real-time inference of locations from tweet text, and deployed the methodology in a system (SAVITR). The proposed methodology performs better than many prior methods, and is much more suitable for real-time deployment. 

We observed several challenges that remain to be solved. For instance, for some geo-tagged tweets, the tweet is posted from a different place as compared to the locations mentioned in the text. 
A common phenomenon is that a tweet posted from a metropolitan city (e.g., Delhi) contains some information about a suburb. How to deal with such 
tweets is application-specific.
Again, there are multiple locations having the same name. Hence disambiguating location names is a major challenge. 
We plan to explore these directions in future.

\bibliographystyle{ACM-Reference-Format}
\bibliography{twitter_references,reference} 

\end{document}